\documentclass[twocolumn,amsmath,amssymb,twoside,preprintnumbers,floatfix,showpacs,pre,superscriptaddress]{revtex4-1}

\usepackage[latin1]{inputenc}
\usepackage[english]{babel}
\usepackage{graphicx}
\usepackage{rotating}
\usepackage[dvips]{epsfig}
\usepackage{amsmath}
\usepackage{amssymb}
\usepackage{bm}
\usepackage{natbib}

\graphicspath{{}{images/}}
\newcommand{\eq}[1]{Eq.\,(\ref{#1})}
\newcommand{\eqns}[2]{Eqns.\,\mbox{(\ref{#1})--(\ref{#2})}}

\newcommand{\fig}[1]{Fig.\,\ref{#1}}
 
\newcommand{\sect}[1]{Sec.\,\ref{#1}}
\newcommand{\be}{\begin{equation}}
\newcommand{\ee}{\end{equation}}
\newcommand{\bea}{\begin{eqnarray}}
\newcommand{\eea}{\end{eqnarray}}
\newcommand{\ba}[1]{\begin{array}{#1}}
\newcommand{\Ref}[1]{Ref.\,\cite{#1}}

\newcommand{\etal}{\emph{~et~al.~}}
\newcommand{\ea}{\end{array}}

\renewcommand{\epsilon}{\varepsilon}
\renewcommand{\theta}{\vartheta}
\renewcommand{\vec}[1]{\mathbf{#1}}

\newcommand{\fracpartial}[2]{\frac{\partial#1}{\partial#2}}

\begin{document}
\title{Implementation of on-site velocity boundary conditions for D3Q19 lattice Boltzmann}
\author{Martin Hecht}
\affiliation{Institute for Computational Physics,
University of Stuttgart,
Pfaffenwaldring 27,
70569 Stuttgart,
Germany}
\author{Jens Harting}
\affiliation{ Department of Applied Physics, 
TU Eindhoven, Den Dolech 2, 
5600 Eindhoven, 
The Netherlands
}
\affiliation{Institute for Computational Physics,
University of Stuttgart,
Pfaffenwaldring 27,
70569 Stuttgart,
Germany}

\date{\today}

\begin{abstract}
\noindent
%\textbf{Abstract.}
On-site boundary conditions are often desired for lattice Boltzmann simulations
of fluid flow in complex geometries such as porous media or microfluidic
devices. The possibility to specify the exact position of the boundary,
independent of other simulation parameters, simplifies the analysis of the
system. For practical applications it should allow to freely specify the
direction of the flux, and it should be straight forward to implement in three
dimensions. Furthermore, especially for parallelized solvers it is of great
advantage if the boundary condition can be applied locally, involving only
information available on the current lattice site. We meet this need by
describing in detail how to transfer the approach suggested by Zou and
He\,\cite{ZouHe} to a D3Q19 lattice. The boundary condition acts locally, is
independent of the details of the relaxation process during collision and
contains no artificial slip. In particular, the case of an on-site no-slip
boundary condition is naturally included. We test the boundary condition in
several setups and confirm that it is capable to accurately model the velocity
field up to second order and does not contain any numerical slip.
\end{abstract}



\maketitle

% ==========================================================================================
\section{Introduction}
% ------------------------------------------------------------------------------------------

\label{sec_Intro}

\noindent
The lattice Boltzmann method (LBM) is a widely used method for the simulation of fluid 
flow\,\cite{Succi01}.  It solves the Boltzmann equation on a discrete lattice and it has been proven that
the Navier Stokes equations can be recovered\,\cite{chen-chen-matthaeus,Higuera89}. The method has 
been successfully applied to the simulation of flow in porous media\,\cite{Kutay06,martys-chen}, 
colloidal suspensions\,\cite{Ladd94, Ladd94b, Ladd01, hartingnaim08}, 
liquid-gas phase transitions and multi-component 
flows\,\cite{shan-chen-93,shan-chen-liq-gas,chen-boghosian-coveney}, 
spinodal decomposition\,\cite{Harting05,chin-coveney}, 
and many more applications.

In spite of the wide range of applications of the LBM 
%Latt and co-workers pointed out in a recent paper\,\cite{Latt08} that 
there is sitll little consensus on how to implement boundary 
conditions in the LBM. For some applications, especially for complex geometries in technical 
applications, rather simple approaches like on-site bounce back\,\cite{Frisch87} rules are 
preferred\,\cite{Ginzburg-Steiner}, and on the other hand quite complex methods to implement
exact boundary conditions have been proposed\,\cite{Skordos93,ginzburg96}. 
A promising approach of velocity boundary conditions by Zou and He\,\cite{ZouHe} for 2D simulations 
has been generalized to 3D with the restriction to the inflow being perpendicular to the boundary 
plane by Kutay\etal\,\cite{Kutay06}.
However, to our knowledge, a generalization to 3D with variable inflow direction has
not yet been presented. Apart from this restriction, in the terms used in \Ref{Kutay06} some 
of the prefactors have to be revised (the correct ones can be found in \Ref{Mattila09}),
but for the application studied by Kutay\etal, the terms used in their work might be appropriate. 
However, especially if the influx direction is not aligned with the computational lattice, 
our slightly more general expressions have to be used. We follow the ideas of Zou and He\,\cite{ZouHe} 
and derive flux boundary conditions with variable influx direction for a D3Q19-lattice\,\cite{qian} 
meaning that in three dimensions the velocity space contains 19 discrete vectors. 
Zou and He\,\cite{ZouHe} have demonstrated a derivation for the D2Q9 model 
and shortly sketched the application to pressure boundaries in a D3Q15i model, where i stands for
an incompressible model of equilibrium distribution functions\,\cite{ZouHou95}. 
Already Zou and He point out that besides the basic idea of applying a bounce back rule
to the non-equilibrium part, a further modification is necessary to achieve the correct 
transversal momentum. A suitable choice for this correction depends on the lattice type. 
Zou and He give an expression for the D3Q15i lattice for the case of pressure boundaries.
In the subsequent publications on D3Q19 lattices\,\cite{Kutay06,Mattila09}, which is one of the most commonly 
used lattice types nowadays, it is assumed that the flux direction is restricted to the 
direction normal to the boundary plane and that this symmetry is also reflected
in the distribution functions on the boundary nodes. In our generalization we drop this 
restriction and consistently derive the transversal momentum corrections. We investigate 
the accuracy of this boundary condition and  highlight the special case of on-site no-slip
boundary conditions included in this approach by simply setting the velocity equal to zero.

Examples for possible applications are microfluidic
devices\,\cite{beskok}, i.e., microscopic channel structures which are specially designed 
to modify a given flow profile by the roughness or wettability of the walls or the 
geometry of the channels. One example for those structures are micromixers\,\cite{Hessel05}.
In general, if simulating such devices, it is not always possible to align all walls with 
the Cartesian planes. In these cases one needs a boundary condition
which is capable to specify the velocity in an arbitrary direction, depending on the orientation
of the channel to be simulated. 

% One might also be interested in the numerical error that occurs due to the staircase-like 
% discretization of the walls depending on the lattice constant used in the simulation. 
% To study this error, one 
% can simulate tilted channels, but the in- and outflux velocity should be specified on the 
% boundary in the direction of the channel, which then does not coincide with one of the main
% lattice directions. In the present paper we use such simulations to test the boundary 
% conditions and to compare it to other approaches.

One might also think of applications in porous media\,\cite{Kutay06}, where flow through 
discretized samples of stones are simulated. On the boundaries of the microscopic pores,
a no-slip condition has to be applied. This is a special case of velocity boundary conditions
with the velocity being zero. Since the channels cannot be aligned with the computational 
lattice, the question raises, how large the error introduced by the discretization is. 
The on-site velocity boundary conditions brought forward in the present paper can be used 
as a replacement of the bounce-back rule for the no-slip condition % and in addition to study 
% the influence of the discretization of the wall, as described above. 
if the velocity is set to zero. In contrast to the usual bounce-back rule the position of the wall
is independent of the BGK relaxation time. This fact is of great advantage when analyzing the 
permeability of a discretized sample.

% The boundary condition we have derived for the D3Q19 lattice may not only be used as 
% in- and outflow condition to study the 
% discretization errors, but it can be a replacement for the bounce back boundary condition. 
% By setting the velocity on the boundary equal to zero one obtains no-slip boundaries as a 
% special case intrinsically included in the velocity boundary condition.
Although the assumption of the fluid velocity being zero on the boundaries does not hold for 
several cases in microfluidics, the no-slip condition is highly important for many cases. 
Therefore, a considerable effort of research has been spent to develop no-slip boundary conditions\cite{inamuro95,Maier96,ginzburg96,Succi01}. Some approaches turned out to
contain an artificial slip length depending on various details of the simulation method,
whereas other attempts involve non-local calculations like the evaluation of a 
velocity gradient to extrapolate the flow field beyond the boundary.
In contrast to that, the boundary condition we propose is of local type and allows 
to specify the velocity on the node exactly with vanishing slip length.

Further, the boundary condition is of great benefit for hybrid  simulations, i.e., 
simulations in which two simulation methods are coupled to simulate 
fluid flow\,\cite{Flekkoy00,delgado03,Delgado-Praprotnik08}. The main goal of such 
hybrid simulations is to save computing time. A computationally
cheap method is applied to simulate the flow on a more coarse-grained level, whereas 
in some regions, where for example interactions on the atomistic level are relevant,
a different simulation method comprising more details is applied. The two simulation
methods are coupled for example by exchange of mass, momentum and energy between the two domains 
via their respective boundary conditions. One can think of different setups:
an LB simulation can be embedded into a finite element based Navier Stokes solver
and resolve one region in more detail. Another example case is that in an LB simulation
one region is resolved even on the molecular level by means of a Molecular Dynamics
simulation. %, which is connected through the boundary conditions to the LB simulation.
Practically, in current hybrid simulations the coupling is implemented within an overlapping region\,\cite{Dupuis07b}
of the two simulations, but it would be a great advance if one could manage simply to couple 
two boundary conditions without any overlap being needed. The possibility to generally 
determine the velocity on a boundary node in an LB simulation is one step towards this goal.

The remainder of this paper is structured as follows: in the following section we 
describe the simulation method in general and introduce our notation of the lattice vectors. 
Then, we shortly review different boundary conditions in the literature in \sect{sec_boundaries}. 
After that, we  derive and discuss the velocity boundary condition for the  D3Q19 lattice
in \sect{sec_onsitevel}. We separately discuss the special case of the 
no-slip condition in \sect{sect_noslip}. Numerical results are presented 
and discussed in \sect{sec_numerics} and finally, we draw a conclusion in the closing section 
of the current paper.

% ==========================================================================================
\section{Simulation method}
% ------------------------------------------------------------------------------------------
\label{sec_method}

\begin{figure}
  \includegraphics[width=\linewidth]{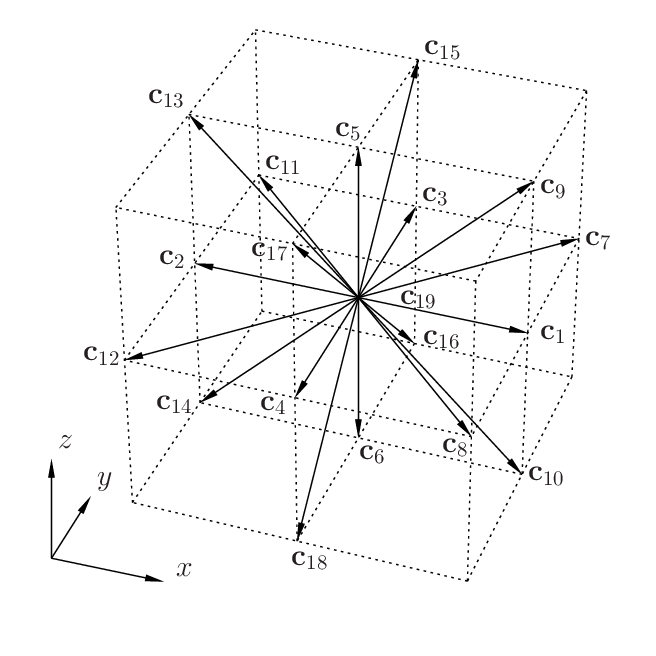}
  \caption{The geometry of the D3Q19 lattice with lattice vectors $\vec{c}_i$ 
    as defined in \eq{eq_deffs}.}
  \label{fig_d3q19}
\end{figure}

\noindent
The lattice Boltzmann method is a numerical method to solve the Boltzmann equation 
\eq{eq_boltzmann} on a discrete lattice. The Boltzmann equation describes the dynamics
of a gas from a microscopic point of view: in a gas, particles, each with velocities 
$\vec{v}_i$, collide with a certain probability and exchange momentum among each other.
For ideal collisions total momentum and energy are conserved in the collisions.
The Boltzmann equation expresses how the probability $f(\vec{x},\vec{v},t)$
of finding a particle with velocity $\vec{v}$ at a position $\vec{x}$ and at 
time $t$ evolves with time:
\be
 \label{eq_boltzmann}
  \vec{v}\cdot\nabla_\vec{x}f + 
  \vec{F}\cdot\nabla_\vec{p}f +
  \frac{\partial f}{\partial t} =
  \hat\Omega(f)\,,
\ee
where $\vec{F}$ denotes an external body force, $\nabla_{\vec{x},\vec{p}}$ the
gradient in position and momentum space, and 
$\hat\Omega(f)$ denotes the collision-operator. Bhatnagar, Gross, 
and Krook\,\cite{BGK} proposed the so-called BGK dynamics, 
where the collision operator $\hat\Omega$ is chosen as a relaxation with a characteristic time $\tau$
to the equilibrium distribution $f^{(eq)}(\vec{v}, \rho)$.
\be
   \label{eq_bgk}
   \hat\Omega(f) =-\frac{1}{\tau}\left(f-f^{(eq)}\right)\,.
\ee
The equilibrium distribution function for athermal models depends
on the local density $\rho(\vec{x}, t)$ and the velocity field $\vec{v}(\vec{x}, t)$.
The lattice Boltzmann method\,\cite{mcnamara-zanetti} discretizes the probability density 
$f$ in space and time. The discrete Boltzmann equation, which is solved by the LBM can be 
rigorously derived from the Boltzmann equation\,\cite{he-luo}. 
The discretization, and especially the analytic expression for the 
equilibrium distribution $f^{(eq)}$ depends on the lattice type. We use a 
D3Q19-lattice which is a very popular lattice type for 3D LB-simulations.
On each lattice site 19 values $f_i({\vec x},t)$ are stored, each of them assigned to a lattice
vector  $\vec{c}_i$. We use the notation that the vectors $\vec{c}_i$ are the $i^{th}$ column 
vector of the matrix 
\begin{widetext}
\be
 \label{eq_deffs}
  \mathbf{M}=\left[\,
  \ba{ccccccccccccccccccc}
    \,1\,&\,-1\,&\, 0\,&\, 0\,&\, 0\,&\, 0\,&\, 1\,&\, 1\,&\, 1\,&\, 1\,&\,-1\,&\,-1\,&\,-1\,&\,-1\,&\, 0\,&\, 0\,&\, 0\,&\, 0\,&\, 0\, \\    
    \,0\,&\, 0\,&\, 1\,&\,-1\,&\, 0\,&\, 0\,&\, 1\,&\,-1\,&\, 0\,&\, 0\,&\, 1\,&\,-1\,&\, 0\,&\, 0\,&\, 1\,&\, 1\,&\,-1\,&\,-1\,&\, 0\, \\
    \,0\,&\, 0\,&\, 0\,&\, 0\,&\, 1\,&\,-1\,&\, 0\,&\, 0\,&\, 1\,&\,-1\,&\, 0\,&\, 0\,&\, 1\,&\,-1\,&\, 1\,&\,-1\,&\, 1\,&\,-1\,&\, 0\, \\
  \ea\,
  \right]\,.
\ee
%\caption{Definition of the lattice vectors $\vec{c}_i$.}
%\label{tab_deff}
\end{widetext}
The geometry is shown in \fig{fig_d3q19} \footnote{Note that some authors, e.g. in \Ref{Kutay06}, use a different notation in which the vectors $\vec{c}_8 - \vec{c}_{12}$ 
are permutated, $\vec{c}_{13}$ and $\vec{c}_{14}$ as well as $\vec{c}_{17}$ and
$\vec{c}_{18}$ are exchanged. Sometimes $\vec{c}_{19}$ is denoted as $\vec{c}_{0}$.}.
The local density at a lattice point can be obtained by
summing up all $f_i$,
\be
  \rho(\vec{x}, t) = \sum\limits_{i=1}^{19}f_i(\vec{x},t)\,,
  \label{eq_rho}
\ee
and the streaming velocity is given by
\be
  \vec{v}(\vec{x},t) = \frac{1}{\rho(\vec{x},t)} \sum\limits_{i=1}^{19}f_i(\vec{x},t)\vec{c}_i\,.
  \label{eq_vel}
\ee
%Note that we 
We express all quantities in lattice units, i.e., time is measured 
in units of update intervals and length is measured in units of the lattice constant.
For practical applications a suitable mapping to physical units based on a 
dimensional analysis has to be applied. 

In the lattice Boltzmann method two steps are performed in  an alternating way:
\begin{enumerate}
  \item The ``streaming step'': propagate each of the distribution functions 
    $f_i$ to the next
    lattice site in the direction of its assigned lattice vector $\vec{c}_i$\,.
  \item The ``collision step'': 
    on each lattice site relax the probability functions $f_i$ 
    towards the equilibrium value $f_i^{(eq)}(\vec{v}, \rho)$. In BGK dynamics 
      this is according to \eq{eq_bgk}.
\end{enumerate}
The equilibrium value $f_i^{(eq)}$ is obtained by discretizing the Boltzmann
distribution. Several expressions of different order have been proposed,
where we use the popular form involving terms in the velocity up to the 
second order\,\cite{qian-dhumieres-lallemand,Tang05,ZouHe,Sbragaglia05,Chen96}
%\bea
\be
     f_i^{(eq)}(\rho, \vec v) = w_i \rho\left[1 + \frac{\vec{c}_i \cdot \vec v}{c_s^2}
      +\frac{(\vec{c}_i \cdot \vec v)^2}{2c_s^4}  - \frac{v^2}{2c_s^2}\right]
%       \right.\nonumber\\
%       & & \qquad\left.+\frac{(\vec{c}_i\cdot \vec v)\left[(\vec{c}_i\cdot \vec v)^2 - 
%       v^2\right] }{2c_s^4}\right]
  \label{eq_feq}
%\eea
\ee
with the lattice speed of sound $c_s = \frac{1}{\sqrt{3}}$ 
for the D3Q19 lattice and the lattice weights  
\be
   w_i = \left\{
   \ba{rcl}
     \frac{2}{36},&\quad &i=1\dots 6 \\
     &\quad &\\
     \frac{1}{36},&\quad &i=7\dots 18 \\
     &\quad &\\
     \frac{12}{36},&\quad &i=19
   \ea\right.
  \label{eq_weights}
\ee
The pressure $p=c_s^2\rho$  turns out to be proportional to the density and the dynamic
shear viscosity is given by\,\cite{Haenel04,Succi01}
\be
  \eta = c_s^2\rho\left( \tau - \frac{1}{2}\right)\,.
  \label{eq_visc}
\ee

\section{Boundary conditions}
\label{sec_boundaries}

\noindent
On the boundary nodes, the distribution function assigned to vectors $\vec{c}_i$
pointing out of the lattice move out of the computational domain in the propagation step, and
the ones assigned to the opposing vectors are undetermined because there are no nodes  
which the distributions could come from. Therefore, on the boundary nodes, special rules have 
to be applied.
 
These boundary conditions can be chosen in various manners. Periodic boundaries 
are realized by propagating the $f_i$ leaving the computational domain on the one boundary 
to the boundary nodes located on the opposite side of the domain. Closed boundaries are commonly 
implemented by a so-called 
mid-grid bounce-back rule\,\cite{Succi01}, which means that the distributions $f_i$ pointing out
of the domain are copied to $f_j$, for which $\vec{c}_j = - \vec{c}_i$, i.e., locally, on each 
lattice site, the undetermined values are filled with the ones which would stream out of
the domain without collision on the boundary node. They enter one time step later into the 
simulation domain again\,\cite{Sukop}.

However, for many questions in fluid dynamics it is required to determine the 
pressure or the velocity field at the boundary.
The first is known as Dirichlet boundary condition, and the latter as Neumann 
boundary condition. In the Neumann case the flux on the boundary of the domain is fixed, 
whereas in the Dirichlet case the pressure is given as a boundary condition.

Zou and He\,\cite{ZouHe} have proposed how to implement Dirichlet and Neumann boundary conditions
on a D2Q9 lattice and shortly sketched how to apply it for a D3Q15i simulation.
Kutay\etal\,\cite{Kutay06} have transferred this proposal to a D3Q19 lattice.
However, their approach is derived under the assumption that the in- and outflow
velocity is always perpendicular to the boundary plane, and oriented along one 
of the main lattice directions ($\vec{c}_i,\ i=1\dots 6$). We generalize this to
inflow with arbitrary direction in \sect{sec_onsitevel}.
%  Although already Zou and He 
% have already presented data from simulations in three dimensions, a description
% how to assign the unknown distributions at the boundary has not yet been reported in 
% the literature. Very recently, Mattila\etal\cite{Mattila09} have used the boundary conditions
% similar as in \Ref{Kutay06}, but with corrected transversal momentum corrections.

Often more elaborated boundary conditions are applied. Chen and co-workers\,\cite{Chen96} 
and Ginzbourg and d'Humi\`{e}res\,\cite{ginzburg96}
suggested extrapolation of the $f_i$ on the first and second layer of the lattice
to the nodes outside the domain. These extrapolated values can be thought 
of as the lattice populations propagating into the domain and arriving on the boundary 
nodes in the next streaming step. Inamuro and co-workers have introduced a counter-slip
to compensate a numerical slip which occurs when applying on-site bounce-back\,\cite{inamuro95}.
Skordos came up with an approach where additional differential equations are solved on 
the boundary nodes to calculate the unknown populations\,\cite{Skordos93}. 
Ansumali and Karlin have developed 
a LB no-slip boundary condition from kinetic theory\,\cite{Ansumali02}, and, more recently, 
d'Orazio\etal\,\cite{dOrazio03} and Tang\etal\,\cite{Tang05} came up with thermal boundary conditions
which also involve an extrapolation scheme and bounce-back with counter-slip respectively.
Ladd and Verberg have developed a boundary condition with a resolution of the position of the wall on a 
sub-grid level, which is  especially required if suspended particles are modeled\,\cite{Ladd01,Ladd01b,Ladd07}. Schiller and D{\"u}nweg\,\cite{Schiller08} use a reduced 
set of distribution functions on the boundary nodes. For their reduced D3Q19 model they derive 
equilibrium distributions and propose a multi relaxation time dynamics and a special collision 
operator on the boundary. 
%It is interesting to note that they find a good numerical accuracy 
%for a Poiseuille flow when they apply their model with a BGK dynamics and
%a bounce-back scheme for the non-equilibrium parts.

Latt\etal have compared and discussed several of these approaches in \Ref{Latt08}. They also 
include the boundary condition by Zou and He\,\cite{ZouHe} in their discussion. As indicated 
by Latt\etal a generalization of the boundary conditions proposed by Zou and He 
is still not provided. However, a general local boundary rule which can be applied in a simple 
way on each node separately, would be desirable. 

We derive such a boundary condition in the following section. Our generalization of the 
velocity boundary condition proposed in\,\Ref{ZouHe}
only involves the distribution functions defined on the local boundary node and allows
by very simple and computationally cheap steps to set the velocity on the node to 
a distinct vector. The desired value is obtained exactly and we cannot detect any 
artifacts like a numerical slip length or bends in the velocity profile.

%==========================================================================================
\section{General on-site velocity boundary condition}
%-----------------------------------------------------------------------------------------
\label{sec_onsitevel}
\noindent As mentioned in the previous section, we extend the boundary
condition by Zou and He\,\cite{ZouHe} to a D3Q19 lattice. We derive the 
boundary condition for the bottom plane ($z = 0$) in detail and give the 
results for the other planes in the appendix. They can be derived following 
the same steps. 

The boundary conditions are
derived by using the set of equations consisting of \eq{eq_rho} and 
the components of \eq{eq_vel}: %, which form the following system of equations
\bea
  \rho v_x &=& f_{1} + f_{7} + f_{8} + f_{9} + f_{10} \nonumber\\
           &&  - (f_{2} + f_{11} + f_{12} + f_{13} + f_{14})\,, \label{eq_vx}\\
  \rho v_y &=& f_{3} + f_{7} + f_{11} + f_{15} + f_{16}  \nonumber\\
           &&  - (f_{4} + f_{8} + f_{12} + f_{17} + f_{18}) \,, \label{eq_vy}\\
  \rho v_z &=& f_{5} + f_{9} + f_{13} + f_{15} + f_{17}  \nonumber\\
           &&  - (f_{6} + f_{10} + f_{14} + f_{16} + f_{18}) \,. \label{eq_vz}
\eea
Due to the continuity relation $\fracpartial{\rho}{t} + \nabla \cdot (\rho \vec{v}) =0$,
we are free to specify only three of the four variables ($\rho$ and 
the three components of $\vec v$) on the boundary. If we fix the tangential 
velocity $v_x, v_y$ on the bottom-layer of the lattice,
and the density to a given value $\rho_0$,
the $z$-component of the inflow velocity $v_z$ can be calculated from \eq{eq_vz}
and \eq{eq_rho},
\bea
    v_z = 1 &-& \frac{1}{\rho_0}\left[f_{1} + f_{2} + f_{3} + f_{4} \right.\nonumber\\
    &+&  f_{7} + f_{8} + f_{11} + f_{12} + f_{19}  \label{eq_rhoin}\\
    &+&\left.  2 ( f_{6} + f_{10} + f_{14} + f_{16} + f_{18})\right]\,,\nonumber
\eea                                      
where the $f_i$ pointing \emph{out of} the system appear with a prefactor of $2$,
and all in-plane components appear with weight $1$. The components pointing into 
the system, $f_{5}, f_{9}, f_{13}, f_{15},$ and $f_{17}$, which are undetermined after 
the streaming step, do not appear at all. With \eq{eq_rhoin} Neumann 
(or pressure) boundary conditions can be applied by specifying $\rho_0$ on the boundary
and using \eq{eq_rhoin} to calculate $v_z$. If \eq{eq_rhoin} is written in the form
\bea
    \rho &=& \frac{1}{1-v_z}\left[ f_{1} + f_{2} + f_{3} + f_{4} \right.\nonumber\\
    &+&  f_{7} + f_{8} + f_{11} + f_{12} + f_{19} \label{eq_uin} \\
    &+&\left.  2 ( f_{6} + f_{10} + f_{14} + f_{16} + f_{18})\right]\,,\nonumber
\eea                                      
all three components of the velocity can be specified and \eq{eq_uin} is used to calculate
the density $\rho$. This is the Dirichlet case, or flux-boundary condition. Again, the 
undetermined populations $f_{5}, f_{9}, f_{13}, f_{15},$ and $f_{17}$ do not enter the 
calculation.
 
We have used two out of four equations (\eqns{eq_vx}{eq_vz} and \eq{eq_rho}), but we still 
have to compute the five $f_i$ pointing into the computing domain. 
Following Zou and He\,\cite{ZouHe} we assume that on the boundary the bounce-back condition 
is still valid for the non-equilibrium part $f_i^*$ of the single particle distribution $f_i$:
\be
 f_i^* = f_i - f_i^{(eq)}\,.
 \label{eq_fnoneq}
\ee
The bounce-back condition in $+z$-direction (in normal direction to the boundary) 
would read as
\be
  f_5^* = f_5 - f_5^{(eq)} = f_6 - f_6^{(eq)} = f_6^* \,,
  \label{eq_bounceback}
\ee
which leads by taking $f_5^{(eq)}$ and $f_6^{(eq)}$ from \eq{eq_fnoneq} to 
\bea
  f_5  &=& f_6 - w_6\rho\left[1  - \frac{v_z}{c_s^2} 
                         +  \frac{v_z^2}{2c_s^4} \right] \nonumber\\
       && \phantom{f_6}+ w_5\rho\left[1  + \frac{v_z}{c_s^2} +  \frac{v_z^2}{2c_s^4}
       \right]\nonumber\\
  &=& f_6 + \frac{2 w_5}{c_s^2}\rho v_z = f_6 + \frac{1}{3}\rho v_z
  \label{eq_f5}\,.
\eea
Here we make use of the fact that the distribution
functions in \eq{eq_feq} are approximated by taking only terms up
to $2^{nd}$ order in $\vec{v}$ into account. However, this approximation could be 
applied directly to \eq{eq_f5} as well. For the derivation of the boundary condition 
it is needed, otherwise the higher order terms would introduce anisotropic effects in 
the boundary rule.%, as explained in the following paragraph.

% Consider a D2Q9 lattice as used in \Ref{ZouHe}. In this case only one 
% additional equation is needed to close the system of equations. However, there are
% three unknown distributions to which one could apply the bounce-back rule in the 2D case.
% Taking higher order terms for  $f_i^{(eq)}$ in \eq{eq_feq} into account would give different 
% solutions for the set of unknown populations if \eq{eq_fnoneq} is applied to different 
% ones of the undetermined lattice populations. If the $f_i^{(eq)}$  are approximated by 
% only terms up to the $2^{nd}$ order in $\vec{v}$, namely in the form of \eq{eq_feq},
% the solutions are consistent for each of the possible choices. This reveals the fact
% that this approach is of second order accuracy in the sense that all information on the 
% first and second moments is obtained from local populations\,\cite{Ginzburg-Steiner}. 
% However, the accuracy of an LB scheme may be influenced by different factors including
% the relaxation time and the grid spacing\,\cite{Holdych04}. 
Generally, in the collision step 
(in \eq{eq_bgk}) higher order terms may be taken into account for the bulk, but for the
boundary conditions a qualitatively different approach, like a higher order extrapolation
scheme, has to be considered when aiming for higher order accuracy.

For the D3Q19 lattice, however, we need two more equations. To keep the
symmetry of the problem, we assume bounce-back of the non-equilibrium part for
all populations $f_i$. This results in four equations, 
\bea
 f_{9}  &=& f_{14} + \frac{2w_{9}}{c_s^2}\rho (v_z + v_x)\,,\label{eq_f9NoNs}\\
 f_{13} &=& f_{10} + \frac{2w_{13}}{c_s^2}\rho (v_z - v_x)\,,\label{eq_f13NoNs}\\
 f_{15} &=& f_{18} + \frac{2w_{15}}{c_s^2}\rho (v_z + v_y)\,,\label{eq_f15NoNs}\\
 f_{17} &=& f_{16} + \frac{2w_{17}}{c_s^2}\rho (v_z - v_y)\,, \label{eq_f17NoNs}
\eea
so that the system of equations is overdetermined. 

Therefore, following the Ansatz by Zou and He\,\cite{ZouHe} for the pressure boundary condition on 
a D3Q15i lattice, we introduce two new variables $N^z_x$ and  $N^z_y$, the transversal momentum 
corrections on the $z$-boundary for distributions propagating in $x$ and $y$-direction, 
respectively. These terms turn out to vanish in equilibrium, but they are non-zero, if 
velocity gradients are present, e.g., when shear flow is imposed by the particular choice of the 
boundary conditions. It turns out that these expressions appear again in the stress tensor.
They reflect the fact that by imposing a transversal velocity component on the boundary,
also stress is imposed to the system. The transversal momentum corrections %turn out to 
involve the populations propagating in the boundary plane in the update rule of the boundary 
condition.
We add the terms to the right hand side and assume that the same expression 
with opposite sign is needed for two of the vectors in the same plane. 
Our Ansatz thus reads as follows:
\bea
 f_{9}  &=& f_{14} + \frac{\rho}{6} (v_z + v_x) - N^z_x \,, \label{eq_f9}\\
 f_{13} &=& f_{10} + \frac{\rho}{6} (v_z - v_x) + N^z_x \,,\label{eq_f13}\\
 f_{15} &=& f_{18} + \frac{\rho}{6} (v_z + v_y) - N^z_y \,,\label{eq_f15}\\
 f_{17} &=& f_{16} + \frac{\rho}{6} (v_z - v_y) + N^z_y \,.\label{eq_f17}
\eea
% In the following it will turn out that the variables $N^z_x$ and  $N^z_y$ 
% vanish if all $f_i$ take their equilibrium value. 
The system of equations (\ref{eq_f9})--(\ref{eq_f17}), together with 
\eq{eq_vx} and \eq{eq_vy} is now a closed system. By \eq{eq_vx} and \eq{eq_vy} 
we specify the tangential components of the velocity $v_x$ and $v_y$, which 
do \emph{not} need to be equal to zero in our approach.
Inserting \eqns{eq_f9}{eq_f17} into \eq{eq_vx} and \eq{eq_vy},
gives an exact solution for $N^z_x$, and  $N^z_y$, respectively:
\bea
  N^z_x &=&  \frac{1}{2}\left[f_{1}+f_{7}+f_{8}-(f_{2}+f_{11}+f_{12})\right] \nonumber\\
      &&   - \frac{1}{3}\rho v_x \,,\label{eq_Nx}\\
  N^z_y &=&  \frac{1}{2}\left[f_{3}+f_{7}+f_{11}-(f_{4}+f_{8}+f_{12})\right] \nonumber\\
      &&   - \frac{1}{3}\rho v_y \label{eq_Ny}
\eea
These %non-equilibrium 
transversal momentum corrections can be inserted into \eqns{eq_f9}{eq_f17}
again and together with \eq{eq_f5} we find explicit expressions for all unknown populations.

Note that in \eq{eq_Nx} and (\ref{eq_Ny}) it is required to sum over all 
in-plane contributions to the velocity in $x$- and $y$-direction 
and the weights are consistent with 
the lattice weights of the $f_i$ appearing in the above expressions. 
As expected, $N^z_x$ and $N^z_y$ vanish
(up to $2^{nd}$ order which is our precision within this derivation),
if we set all $f_i$ to their equilibrium value. 
The results for the other planes are given in the appendix. 

A general form for all boundary planes can be written down by introducing the 
normal vector on the boundary $\vec{n}$, the tangential vectors 
$\vec{t}_i=\vec{c}_i-(\vec{c}_i\cdot\vec{n})\vec{n}$, and the notation 
$f_{-i}$ denoting the direction to which a population is bounced back $\vec{c}_{-i} = - \vec{c}_{i}$.
From the populations $f_i$ assigned to a direction $\vec{c}_i$ pointing into the wall, 
the new populations $f_{-i}$ in opposite direction can be calculated as
\be
  f_{-i} = f_{i} - \frac{\rho}{6}\,\vec{c}_i\cdot\vec{v} - \frac{\rho}{3}\,\vec{t}_i\cdot\vec{v}
    +  \frac{1}{2}  \sum\limits_{j=1}^{19}f_j\,\left(\vec{t}_i\cdot\vec{c}_j\right)\, 
    \left(1-\left|\vec{c_j}\cdot\vec{n}\right|\right)\,.
 \label{eq_bbshort}
\ee

Due to the particular choice of \eqns{eq_f9}{eq_Ny} or \eq{eq_bbshort} respectively,
it is possible to specify the 
velocity to an exact value on the lattice site. The rules presented here are
independent on the relaxation rate in the collision step, since all calculations
involve only the known values of the $f_i$ and equillibrium functions. 
Relaxation is calculated separately after all unknown $f_i$ are calculated and 
the macroscopic velocity and density are preserved during collision. 
There are no restrictions on the orientation of the inflow direction. 
Furthermore, all calculations are local on each lattice site. 
Apart from using only terms of first and second order in $\vec{v}$ for the 
equilibrium distributions $f_i^{(eq)}$ in \eq{eq_feq} no approximations are made.
% By introducing the  transversal momentum corrections  $N^z_x$ and $N^z_y$ and thus 
% keeping the freedom to solve the above equations exactly, no further approximations 
% which would reduce the order of the boundary condition are necessary. 
% The above equations are solved exactly. 
Therefore, we have derived a way to implement explicit local on-site boundary conditions 
which model the fluid field up to second order in the velocity.

A similar scheme as ours has been proposed by Halliday\etal\,\cite{Halliday} for a 
D2Q9 lattice. These authors construct the unknown distributions locally on each 
lattice site starting from a Chapman-Enskoog analysis. During their derivation
they have to choose a set of variables they consider as free variables. This is 
similar to the approach of introducing the transversal momentum corrections in order 
to be able to solve the system of equations. Halliday\etal find results for the
unknown populations which involve the components of the strain rate tensor calculated
from the known populations. From this point of view it might be
possible to apply a scheme similar to the one proposed by Halliday\etal  to a D3Q19
lattice. However, in three dimensions the systems of equations, in the generality 
of \Ref{Halliday}, might become difficult to handle. 

Special care has to be taken when connecting the in- and  outflux boundary 
conditions at the corners and edges of the simulation domain with other types of 
boundary conditions that are applied on other boundary planes.
% We apply mid-grid bounce back or periodic boundaries in $x$- and $y$-direction
% in conjunction with the streaming step. Then we treat in- and outflux on the 
% $z$-boundary-planes, so that afterwards all $f_i$ are defined, 
% when performing the collision finally. Note that 
If no-slip boundary conditions are assumed on the $x$- and $y$-boundary, 
% connected to this new boundary condition, 
one has to take care that the influx velocity tends to zero at the edges. 
% $N^z_x$ and $N^z_y$ then vanish at the edges if calculated after applying the bounce 
% back rule and if $\vec{v}=0$ is assumed. Thus, \eqns{eq_f9}{eq_f17} also reduce
% to a simple bounce-back rule. For moving boundaries the concept of buried links
% \,\cite{Maier96} can be applied to the edge nodes in our case, as well.
We discuss the special case of no-slip boundaries as a subset of velocity boundaries 
in the following section.

\section{On-site no-slip boundary condition}
\label{sect_noslip}

The on-site velocity boundary condition proposed in this paper includes an important
special case: setting the velocity $\vec{v}=0$ results in a no-slip-boundary 
for non-moving boundaries. 
Therefore, this boundary condition can also be used as a replacement of the mid-grid bounce back rule.  However, even more generally, moving boundaries, e.g., moving 
shear plates, can be implemented by imposing the wall velocity $\vec{v}$ on the boundary nodes. 
The position of the wall is \emph{exactly} on the lattice nodes. This is in contrast to most no-slip boundaries
proposed in the literature, where the wall position is assumed at half the distance between
two nodes. However, in many of those approaches the exact position of the wall depends 
on the BGK relaxation time. This is not the case for our approach. %the boundary condition we present here. 

One of the pillars of the LBM is local mass conservation\,\cite{Latt08},  which should be 
fulfilled not only in the bulk, but also on closed boundaries. However, some 
extrapolation schemes may be less accurate in this point\,\cite{Schiller08}, whereas
for our on-site approach mass conservation is strictly fulfilled at the closed walls.

The  transversal momentum corrections similar to those given in \eq{eq_Nx} and \eq{eq_Ny} are given
in the appendix for each coordinate plane, and both velocity components in 
each of those planes. They are corrections to the on-site bounce back rule. With these 
corrections taken into account, the velocity is exactly zero on the node.
For edge nodes with both boundaries being implemented as described here, we suggest
to first apply the bounce back rule for all $f_i$ pointing out of the computational
domain and then to calculate the  transversal momentum correction. On an edge node only one 
tangential vector along the edge can be used for ensuring no-slip. 

Consider for example the edge between the $xy$-plane and the $yz$-plane, where the 
$y$-axis forms the edge.
Contributions in the boundary planes known after bounce back are $f_1$, $f_7$, and $f_8$
in the $xy$-plane and $f_5$, $f_{15}$, and $f_{17}$ in the $yz$-plane. 
However, to ensure no-slip one can define
\be
  N_y^{xz} =  \frac{1}{4}\left[f_{3}-f_{4}\right] \label{eq_Nxzy}\,.
\ee
The correction to the distributions $f_i$ with $i = 7, 8, 15$, and $17$ then is
$N_y^{xz}\,\vec{c}_i\cdot\vec{t}_i$ which has to be added to the distributions.
The prefactor in \eq{eq_Nxzy} takes into account that the remaining slip velocity after bounce
back is distributed among four populations obtained from the bounce back rule. 
Similar expressions can be written down for each edge.
A general expression for the modified bounce back rule is
\be
  f_{-i} = f_i  -  \frac{1}{4} 
  \sum\limits_{j=1}^{19} f_j\,\left(\vec{t}_i\cdot\vec{c}_j\right)\, 
    \left(1-\left|\vec{c_j}\cdot\vec{n}^{(1)}\right|\right)\,
    \left(1-\left|\vec{c_j}\cdot\vec{n}^{(2)}\right|\right)\,,
    \label{eq_edge}
\ee
where $\vec{n}^{(1)}$ and $\vec{n}^{(2)}$ denote the two normal vectors on the two boundary 
planes meeting at the edge under consideration.

On the edges and corners, apart from the incoming populations, there are so-called
``buried links''\,\cite{Maier96}, i.e., lattice vectors $\vec{c}_i$ for which the 
opposing lattice vector $\vec{c}_{-i}$ points out of the domain, as well.
The two lattice vectors $\vec{c}^{(1,2)} = \pm\left(\vec{n}^{(1)}-\vec{n}^{(2)}\right)$ 
make up the buried link on an edge node. In the following, lattice vectors with subscript, 
$\vec{c}_i$, denote distinct vectors as defined in \eq{eq_deffs}, whereas vectors with 
superscript, $\vec{c}^{(i)}$, denote vectors which belong to the buried links, and which 
depend on the normal vectors on the individual boundary planes. The distribution functions
assigned to the buried links have to be assigned separately. We choose them such that they 
contribute to the same density according to their lattice weight:
\bea
  f^{(1,2)} &= \frac{1}{22}\sum\limits_{i=1}^{18}f_i &  \left\{1 - \label{eq_buried}
     \left(1-\left|\vec{c}_i\cdot\left[\vec{n}^{(1)}\times\vec{n}^{(2)}\right]\right|\right) \right. \\
   && \left. 
    \cdot \left(1-\left|\vec{c}_i\cdot\left[\frac{\vec{n}^{(1)}+\vec{n}^{(2)}}{\left|\vec{c}_i\right|^2}\right]\right|\right)\right\}\,.
     \nonumber
\eea
Similarly, the distribution on the resting node is chosen as
$f_{19}=\frac{w_{19}}{w_7} f^{(1,2)}= 12\, f^{(1,2)} $. The weights are always determined by the 
number of $f_i$ which contribute to the sum and their respective lattice weights $w_i$ according
to \eq{eq_weights}. In \eq{eq_buried} six $f_i$ with lattice weight $\frac{1}{18}$ and ten $f_i$
with weight $\frac{1}{36}$ contribute, which makes up an overall contribution of $\frac{22}{36}$.
To reduce this to the desired lattice weight, we have to divide by 22, and to obtain a value for the
resting node, we multiply by 12, because of the twelve times larger lattice weight of the resting node.

At the edges surrounding the in- and outlet planes, on the other hand, one needs either
pressure or velocity boundary conditions, depending on the boundary type used for in- or outlet.
For velocity boundaries one has to take care that the velocity profile decays to zero, 
so the no-slip boundary condition just described can be used on all edges.
For pressure boundaries one prescribes a density $\rho=\rho_0$ which we can be used to 
calculate the distribution assigned to the buried link:
\be
  {\tilde f}^{(1,2)} = \frac{\rho_0 - 22 f^{(1,2)}}{14}\,\,,
    \label{eq_buried_rho}
\ee
where $f_{19}$ is then calculated as $f_{19} = 12\,{\tilde f}^{(1,2)}$\,.

According to Maier\etal\,\cite{Maier96} no-slip boundaries cannot be enforced on convex boundary 
nodes. However, slip along the edge can be reduced by correcting all distributions $f_i$ traveling
into the interior of the system by
\be
   N_{i} = \frac{1}{4}\,\vec{c}_i\cdot\left(\vec{n}^{(1)}\times\vec{n}^{(2)}\right)\,\sum\limits_{j=1}^{19}f_j\,\vec{c}_j\cdot\left(\vec{n}^{(1)}\times\vec{n}^{(2)}\right)\,,
    \label{eq_edge2}
\ee
which follows the same idea as \eq{eq_edge} and \eqns{eq_f9}{eq_f17}: momentum in a direction 
parallel to the surface, which would remain on a node after applying the boundary rule, 
is removed by modifying those populations that will afterwards propagate back into the 
bulk of the system. In principle, one could split \eq{eq_edge} into two steps: first, apply
bounce back for all populations leaving the system, and then correct the populations traveling
away from the edge by the term given in \eq{eq_edge2}. For convex edges these are the populations
traveling into the bulk, and for concave edges they propagate in the boundary planes.
This opens a possibility to implement all rules in a single procedure, for which the normal vector
$\vec{n}$ is stored on each lattice site by an integer number. The vector is obtained from  
the matrix $\mathbf{M}$ defined in \eq{eq_deffs}. For values between 1 and 6
\eq{eq_bbshort} is applied, for values between 7 and 18 either \eq{eq_edge2} applies or depending on 
the values stored on the neighboring nodes, the bounce-back rules corrected according to \eq{eq_edge} 
may be applied instead. This information, which expresses if the edge is concave or convex,
can be obtained once, when the lattice is generated and may be stored in the sign of the lattice
vector index. The normal vector points into the bulk and indicates the direction of the symmetry plane
on the edge nodes.

Finally, on the corner nodes one can define three normal vectors on the boundary planes 
meeting there, $\vec{n}^{(1)}$, $\vec{n}^{(2)}$, and  $\vec{n}^{(3)}$. Similar to 
the buried links, there is a complete plane in which six vectors are located, that
only couple to the simulation in the collision step. The buried vectors $\vec{c}^{(1\dots 6)}$ 
are the ones for which 
$\vec{c}^{(i)}\cdot\left( \vec{n}^{(1)} +\vec{n}^{(2)}+\vec{n}^{(3)}\right)=0$\,.
Since the normal vector on this plane is not contained in the set of vectors for the D3Q19 lattice
additional indices are needed to mark the corner nodes.
After bouncing back the known $f_i$, the distributions assigned to buried vectors are 
set to 
\be
  f^{(1\dots 6)} = \frac{1}{18}\sum\limits_{i=1}^{18}f_i
     \left[\left|\vec{c}_i\cdot
     \left(\frac{\vec{n}^{(1)}+\vec{n}^{(2)}+\vec{n}^{(3)}}{\left|\vec{c}_i\right|^2}\right)\right|\right]
    \label{eq_corner}
\ee
if velocity boundaries are applied, 
or to ${\tilde f}^{(1\dots 6)} = \frac{\rho_0}{18} - f^{(1\dots 6)}$ if pressure boundaries
are chosen. $f_{19}$ is set to $12\,f^{(1\dots 6)}$ or $12\,{\tilde f}^{(1\dots 6)}$, 
respectively. A correction similar to \eq{eq_edge} is not necessary on the corner nodes\footnote{Values of 19 onwards as vector index can be used to distinguish the 
different corner nodes}. 

In complex geometries there are points in which edges (convex or concave ones) meet planes
which are oriented perpendicular to the direction of the edge. There, we propose to 
use bounce back for those populations which would leave the computational domain and to 
assign an appropriate value to the resting node, similar as described for the corner nodes.
There are no buried links, because those links are located inside the boundary plane which the edge 
connects to, so the resting node must be set to $f_{19}=\frac{12}{36}\sum_{i=1}^{18}f_i$\,.
In total there are 6 planes and 4 possible orientations of the edges, each of them either convex 
or concave, making up 48 more cases. However, since there are no buried links and a momentum
correction is not necessary either, the rules can be implemented easily in only a few lines of code.
In the following section we show the results of tests of the boundary condition
in simple geometries like Poiseuille flow between two plates, where the exact solution is known. 
As an example for more complex geometries we simulate the flow through a rectangular channel,
where also edges and corners are involved.

%==========================================================================================
\section{Numerical Results}
%-----------------------------------------------------------------------------------------
\label{sec_numerics}
\noindent
%We use the three dimensional lattice Boltzmann code LB3D\,\cite{Harting05}
%for our simulations. 
%The code has been extended to use several kinds of boundary conditions,
%not only the type we have derived here, but also the approach suggested by 
%Kutay\etal\cite{Kutay06}, which as the authors point out, is restricted to 
%in- and outflow velocities aligned with one of the main lattice directions.
%
We test our boundary condition by simulating a Poiseuille flow through a tilted channel.
The size of the computational domain is $64 \times 8 \times 128$ LB nodes, where the channel has a width
of 20 nodes and is tilted by an angle $\alpha = \arctan(\frac{40}{127}) \approx 17.48^{\circ}$. 
This angle is chosen such that both ends of the channel intersect the $xy$-plane at the top 
and the bottom of the computational domain and that there are two lattice sites of wall at the 
left and at the right of the channel at the bottom and the top plane respectively. 
The flow through our test channel is simulated in three dimensions. However, for convenience, the 
$y$-direction is periodic.The walls (only in this test) are implemented as simple bounce-back nodes. 
Here we apply the boundary conditions derived in \sect{sec_onsitevel} as in- and outflux conditions and
compare them to other implementations.

We choose this simple test because the analytical solution for the flow field is known 
and so we can estimate the numerical error.  Usually, one would avoid to have walls
not aligned with the computational lattice because of the staircase like discretization 
of the walls, which brings an additional discretization error into the simulation. 
This discretization error can be avoided by simply aligning the channel with the 
computational lattice. However, if more complex structures, like, e.g., Y-channels 
for applications in microfluidics  are simulated, it may happen that always at least one 
channel is not aligned with one of the Cartesian directions. A technical workaround, if appropriate 
boundary conditions are missing, is to simulate a very long channel so that in the first section 
of the channel, the fluid can relax to a steady flow profile, and only afterwards enters the actual 
simulation domain. However, this causes the computational effort to increase substantially.
%, just to obtain a steady flow in a given direction. 

% \begin{figure}
%   \includegraphics[width=\linewidth]{Fig2}
%   \caption{Velocity field: simulated flow through a tilted channel (a), 
%       expected flow field (Poiseuille flow) discretized on the same lattice (b),
%       difference field (c). The velocity is color coded from blue (slow) to
%       red (fast) flow. The velocity vectors are scaled up for drawing by a factor of 200 
%       for the simulated and calculated flow and by 1000 for the difference field.}
%   \label{fig_vfieldnew}
% \end{figure}

% In \fig{fig_vfieldnew}\,a) we show a 2D cross section in the $xz$-plane through the flow field 
% after 20000 time steps. 
% The size of the computational domain is $64 \times 8 \times 128$ LB nodes, where the channel has a width
% of 20 nodes and is tilted by an angle $\alpha = \arctan(\frac{40}{127}) \approx 17.48^{\circ}$. This angle is chosen such that
% both ends of the channel intersect the $xy$-plane at the top and the bottom of the computational
% domain and that there are two lattice sites of wall at the left and at the right of the 
% channel at the bottom and the top plane respectively. 
% The colors in  \fig{fig_vfieldnew} are assigned the absolute value of the velocity. The 
% length of the vectors (in lattice units) is enlarged by a factor of 200 for drawing.
% For drawing of the theoretical field the same color scale as for the simulation results is used. 
% Note that due to the discretization, 
% the steps  in the staircase-like wall are not always of the same height.

Knowing the width of 
the channel, the center of the in- and outlet, and a given velocity $\vec{v}_0$ on the center line, 
one can calculate a Poiseuille flow field inside the channel,
\be
    \vec{v}^P(\vec{x})=\vec{v}_0\left[ 1 - 
       \left(\frac{x-x_0-\gamma(z-z_0)}{\Delta x}\right)^2  \right]  \,,\label{eq_pois}
\ee

%which is shown in \fig{fig_vfieldnew}\,b) after the values
%of the flow field have been discretized on the same lattice as the simulation was performed.
%After 5000 time steps
%a steady flow field is reached. However, to be sure that the simulations have converged, 
%we simulate 20000 time steps until we evaluate data.
%In \eq{eq_pois} 
where $x_0$ and $z_0$ denote the center of the simulation space, $\Delta x$ is the 
half width of the channel measured along the $x$-direction, and $\gamma$ denotes the increment 
due to the tilting angle, which is related to the components of the velocity by 
$\gamma = \tan\alpha = \frac{v_z}{v_x} = \frac{40}{127}$.

We simulate flow through such a channel and apply different in- and outlet boundary conditions.
We use a relaxation time $\tau = 1$ in all simulations presented here. 
However, we have checked that the results do not depend on this particular choice.
After 5000 time steps a steady flow field is reached. However, to be sure that the simulations 
have converged, we simulate 20000 time steps until we evaluate data.

\begin{figure}
  \includegraphics[width=\linewidth]{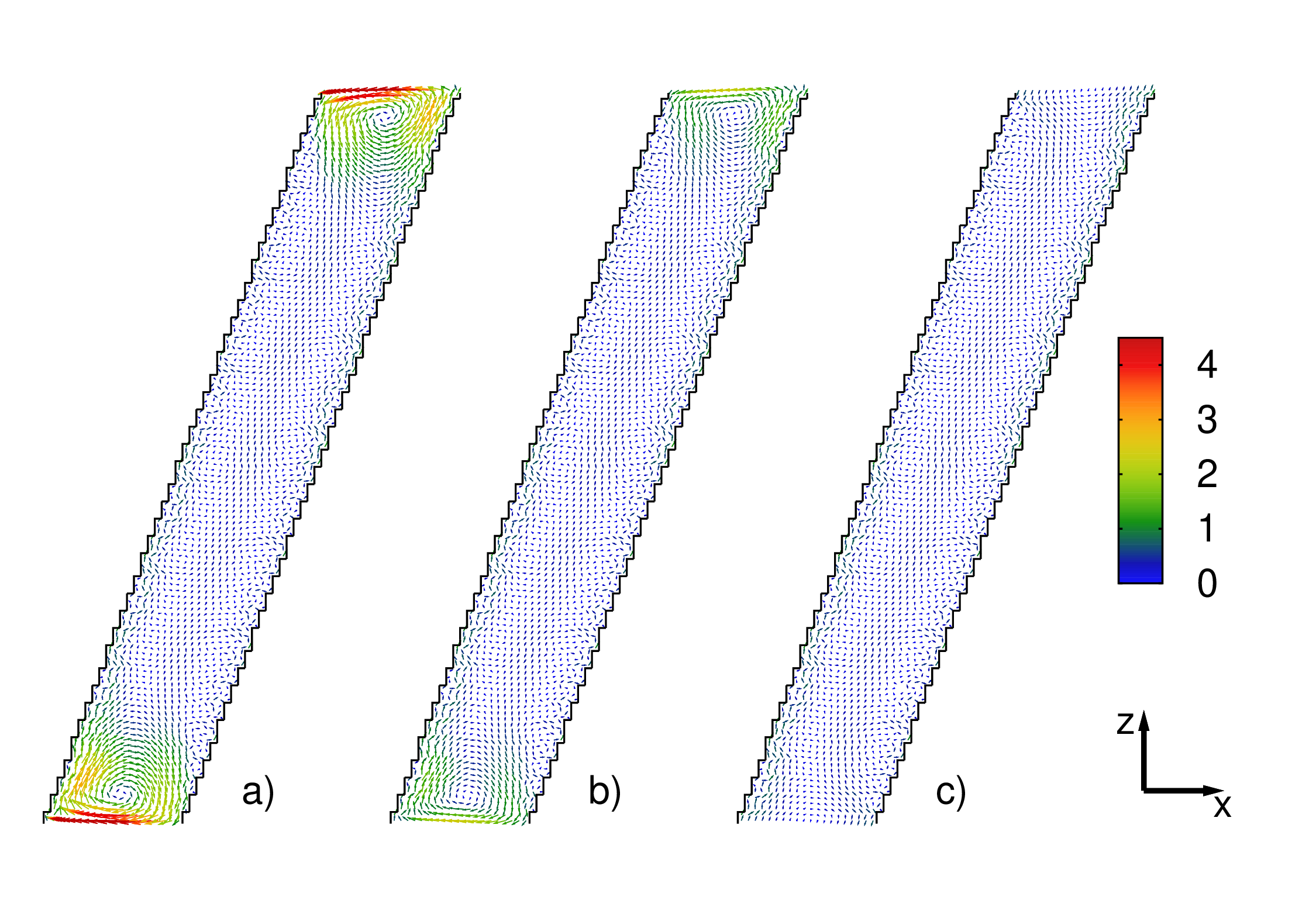}
  \caption{Velocity difference fields for different approaches: In- and outflow 
    velocity constraint to the direction perpendicular to the boundary plane (a),
    In- and outflow velocity tilted and of parabolic shape as in the analytic solution,
    but with $N^z_x$ and $N^z_y$ set equal to zero (b), the boundary conditions
    derived in \sect{sec_onsitevel} with the correct choice for $N^z_x$ and $N^z_y$.
     difference field (c). 
     The velocity difference vectors are scaled for drawing by a factor of 1500 and 
    the absolute value of the velocity difference is color coded from blue (small) 
    to red (large).}
  \label{fig_vdifffields}
\end{figure}

To visualize the difference between simulation and theoretical prediction
we subtract the velocity on each lattice node and draw the resulting vector field in \fig{fig_vdifffields}. 
The value of the velocity is scaled by a factor of 1500 for drawing the arrows. The colors are 
assigned the absolute value of the velocity \emph{after} scaling the difference field.
% For drawing the difference field we use a scaling factor of 1000 
% and the same color coding as in \fig{fig_vfieldnew}\,a) and b) is applied \emph{after} scaling for 
% the original fields and for the 
% difference field. Blue and green color denotes small values and yellow or red color indicates 
% large values.

% The difference in \fig{fig_vfieldnew} can be mostly ascribed to the discretization
% on the lattice. Each single step of the wall discretized to individual steps can be found in 
% the flow profile. However, at the in- and outflux boundary no additional artifacts can be seen,
% which demonstrates the strength of our boundary condition. The fact that the staircase like discretization dominates the numerical error in this test motivates to further 
% analyze the quality of the boundary condition, but first we want to compare the results
% for the tilted channel with results obtained using different in- and outflow conditions.

%In \fig{fig_vdifffields} we show the difference fields for further selected boundary 
%conditions. 
In \fig{fig_vdifffields}\,a) we apply the boundary condition used by 
Kutay\etal\,\cite{Kutay06} to a case, where the restriction of the in- and outflow 
velocity parallel to the $z$-direction introduces an error in the region close to the boundary. 
Note that in \Ref{Kutay06}, apart from assuming the velocity perpendicular to the boundary, 
the authors underestimate the transversal components, which may be of no importance
in this case. We use the correct coefficients as presented very recently in \Ref{Mattila09}, 
but keep the restriction to the inflow perpendicular to the boundary, 
which has a much larger influence on 
the flow field. Not only the first and second layer of nodes close to the boundary
are affected, but the boundary condition introduces vortices which have approximately the 
size of the diameter of the channel.
Therefore, the first step to generalize the boundary condition from \Ref{Kutay06}
to a case where the in- and outflow velocity has an arbitrary orientation, is to 
use \eqns{eq_f9NoNs}{eq_f17NoNs}, as used in the simulation for which the result is 
shown in \fig{fig_vdifffields}\,b). It is obvious that this boundary condition
still introduces vortices close to the in- and outflow. The strength of the vortices is smaller 
compared to the case shown in \fig{fig_vdifffields}\,a). However, the size of the vortices is 
comparable to the width of the channel here as well. 
%Introducing the  transversal momentum corrections 
%$N^z_x$ and $N^z_y$ in \eqns{eq_f9}{eq_f17} on the boundary finally avoids the vortices as shown above.
The value of the tangential velocity on the boundary nodes differs from the value one inserts 
into the equations. By introducing the transversal momentum corrections $N^z_x$ and $N^z_y$ 
in \eqns{eq_f9}{eq_f17}, the vortices disappear and the velocity takes exactly the value 
one specifies with \eqns{eq_vx}{eq_vz} as one can see in \fig{fig_vdifffields}\,c).
The remaining differencefield can be mostly ascribed to the discretization
on the lattice. Each single step of the wall discretized to individual steps can be found in 
the flow profile. However, at the in- and outflux boundary no additional artifacts can be seen,
which demonstrates the strength of our boundary condition. The velocity on the boundary nodes
takes exactly the value which we specify, and therefore, no vortices are generated.

The transversal momentum corrections $N^z_x$ and $N^z_y$ could also be understood in terms of a 
counter-slip similar to the approach of Inamuro\etal\,\cite{inamuro95}, but the Ansatz how to 
obtain the unknown populations $f_i$ is different: we assume a bounce back rule for the 
non-equilibrium part of the distributions and end up with a linear correction to the reflected 
populations, whereas the authors of \Ref{inamuro95} construct the unknown distributions 
based on kinetic theory where the correction appears not on the level of the distribution
functions but as a counter-slip on the level of the wall velocity. The values for the density 
and the velocity inserted into the equilibrium distributions in Inamuro's method are different
from the ones used for bulk nodes. In our approach, however, the boundary nodes are treated 
similar as the bulk nodes: the velocity on the boundary node can be calculated by inserting
\eq{eq_Nx} and \eq{eq_Ny} into \eqns{eq_f9}{eq_f17} and the obtained distributions $f_i$ together with
the one from \eq{eq_f5} and the density from \eq{eq_uin} into \eq{eq_vel}. 
It turns out that the velocity calculated from \eq{eq_vel} is exactly the one which 
is imposed at the boundary node by \eqns{eq_f9}{eq_f17}.

%The quadratic terms 
%of the counter-slip velocity in Inamuro's approach cancel out each other when its influence 
%on the tangential velocity is calculated. Therefore, the accuracy of both ways to specify 
%no-slip boundary conditions is comparable. It might even be possible to deduce an expression 
%which relates the counter-slip in a generalization to a D3Q19 lattice with the  transversal momentum 
%corrections we propose. 

% In \fig{fig_vdifffields}\,c) we show 
% a case where the boundary conditions affect the whole simulation domain. Here, pressure 
% boundaries are applied and the velocity on the boundary is restricted to the $z$-direction.
% This case just demonstrates that obtaining a more or less parabolic profile which in first 
% approximation looks fine, is by far not sufficient and that an appropriate choice of the 
% boundary conditions is essential for obtaining meaningful simulations.

In all simulations we kept the tilting angle of the channel constant, because the error of 
our boundary conditon is angle independent.
We can quantify the quality of the boundary conditions by computing the ratio of the absolute 
value of the difference field and the calculated velocity field on each node. The obtained 
values are averaged over the first twenty layers of LB nodes from the boundary. 
\be
  \xi = \int\limits_{V} \frac{\left|\vec{v}(\vec{x})-\vec{v}^P(\vec{x})\right|%
                             }{\left|\vec{v}^P(\vec{x})\right|}\mathrm{d}V\,,
 \label{eq_deferr}
\ee
Where the volume $V$ contains those layers of lattice nodes, which are at most a distance 
of the channel width apart from the boundary of the computational domain. 
This captures approximately the 
vortices and provides a measure for the quality of the boundary condition. The results for the 
different cases shown in \fig{fig_vdifffields} are listed in the 
following tabular:\\

\begin{tabular}{|l|l|}
\hline\quad
{\bf Boundary condition} & {\bf relative error $\xi$ } \quad\\\hline\quad
on-site velocity (\fig{fig_vdifffields}\,c) \quad&\quad 0.0996 \quad\\\hline\quad
$N^z_x$ and $N^z_y$ set to zero (\fig{fig_vdifffields}\,b) \quad&\quad 0.126 \quad\\\hline\quad
$v_x = v_y = 0$ (\fig{fig_vdifffields}\,a) \quad&\quad 0.175 \quad\\\hline%\quad
%pressure boundary (\fig{fig_vdifffields}\,c) \quad&\quad  0.238 \quad\\\hline
\end{tabular}\\[1ex]

Good agreement with the expected Poiseuille flow profile (\eq{eq_pois}) is reflected in small
relative errors. Large numbers indicate deviations in the area, where the fluid fields are compared. 
We ascribe the 
remaining deviations to the discretization error of the wall and the accompanying uncertainty in
the exact wall position in the present case of the staircase like discretization.
We check this by increasing the resolution of the simulation by a factor of two. 
As we expect, the numerical error due to the staircase-like discretization 
decreases roughly by a factor of two to $\xi= 0.051$. This shows that the staircase-like
discretization introduces a first order error. Therefore, we need further investigations
to see the second order accuracy of the in- and outflux boundary condition.

As another test for the flux boundary condtion we simulate a straight 
channel aligned parallel to the computational lattice, again in a $64\times 8\times 128$-domain 
with the same boundary conditions as for the inclined channel. The remaining
relative error decreases to $0.00235$, which is typical for Poiseuille flow simulations
at this resolution in combination with a mid-grid bounce back rule on the boundary.

We can further measure the quality of our boundary condition in a shear simulation.
On a $32^3$ lattice we apply periodic boundaries in $x$- and $y$-direction and impose a shear 
velocity of $v_x=\pm 0.02$ with opposite sign on the top and bottom plane. We obtain a linear
flow profile within floating point precision. % errorbars of $\pm 7.5\times 10^{-6}\%$ of the desired shear velocity on the 
%boundary. The error fluctuates throughout the simulated volume and can be ascribed to rounding
%errors within the precision of the representation of the floating point numbers on the computer.
There are no notable jumps between the first and second layer of LB nodes, which %again 
confirms that the strain rate tensor ${\bf \Pi}$ is set up correctly on the boundary nodes. 
%Taking higher order terms in \eq{eq_feq} into account for the collision step does 
%not affect the simulation results.

\begin{figure}
  \includegraphics[width=\linewidth]{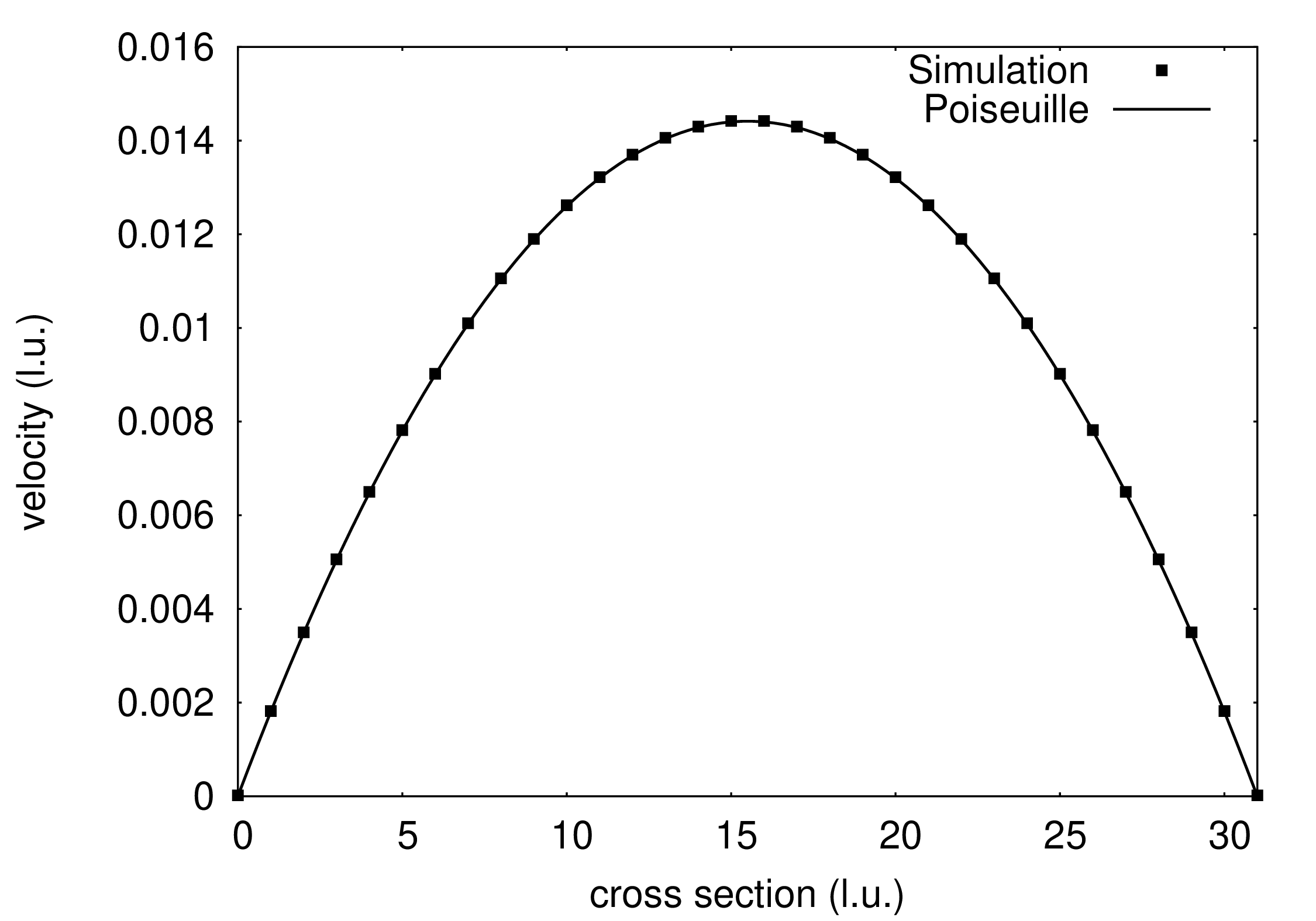}
  \caption{Poiseuille flow between two parallel walls driven by a body force. The simulation
    data are averaged in each lattice plane parallel to the walls and agree up to floating point 
    precision with the calculated Poiseuille profile.}
  \label{fig_poiseuilletest}
\end{figure}

In a next step we simulate Poiseuille flow again, but this time we use a $32^3$
lattice with periodic boundaries in $y$- and $z$- direction. We apply a body force\,\cite{Harting05,bib:guo-zheng-shi-02}
by adding a force term 
\be
  \Delta \vec{v} = \frac{\tau \vec{F}}{\rho} 
   \label{eq_gravity}
\ee
to the velocity in \eq{eq_feq} in the whole simulation volume. The Poiseuille profile we 
expect is of the form 
\be
  \vec{v} = \frac{\vec{F}}{2\eta}\left(1 - \left(\frac{x-x_0}{\Delta x}\right)^2\right)\,,
  \label{eq_poiseuille}
\ee
where the viscosity is given by \eq{eq_visc}.
%A fit to the simulation data converges within 
%the floating point precision exactly to this viscosity, where the velocity is exactly zero 
%on the boundary nodes, even after the next collision step after applying the boundary condition.
The velocity profile found in the simulation together with the expected Poiseuille profile
is plotted in \fig{fig_poiseuilletest}. The parabola contains no fit parameters.
The velocity is exactly zero on the boundary nodes, whereas with a simple bounce-back a numerical
slip can be observed, which results in a velocity of $4\times 10^{-5}$ for the same 
simulation setup without using the transversal momentum corrections $N^z_x$ and $N^z_y$. 
%The parabola was fitted with three parameters to the
%simulation data and it turned out that all three parameters converge within floating point 
%precision to the expected value. 
We have carried out this test with $\tau = 1$, but the data presented in \fig{fig_poiseuilletest}
is obtained with $\tau = 2$ to ensure that our boundary conditions are not restricted to the 
special case of $\tau = 1$. Apart from the influence of $\tau$ on the viscosity (\eq{eq_visc}),
our simulation results are not affected by the relaxation time. In particular, we do not see any 
$\tau$-dependent (numerical) slip.

\begin{figure} 
  a)\hspace{-3ex}\includegraphics[width=0.85\linewidth]{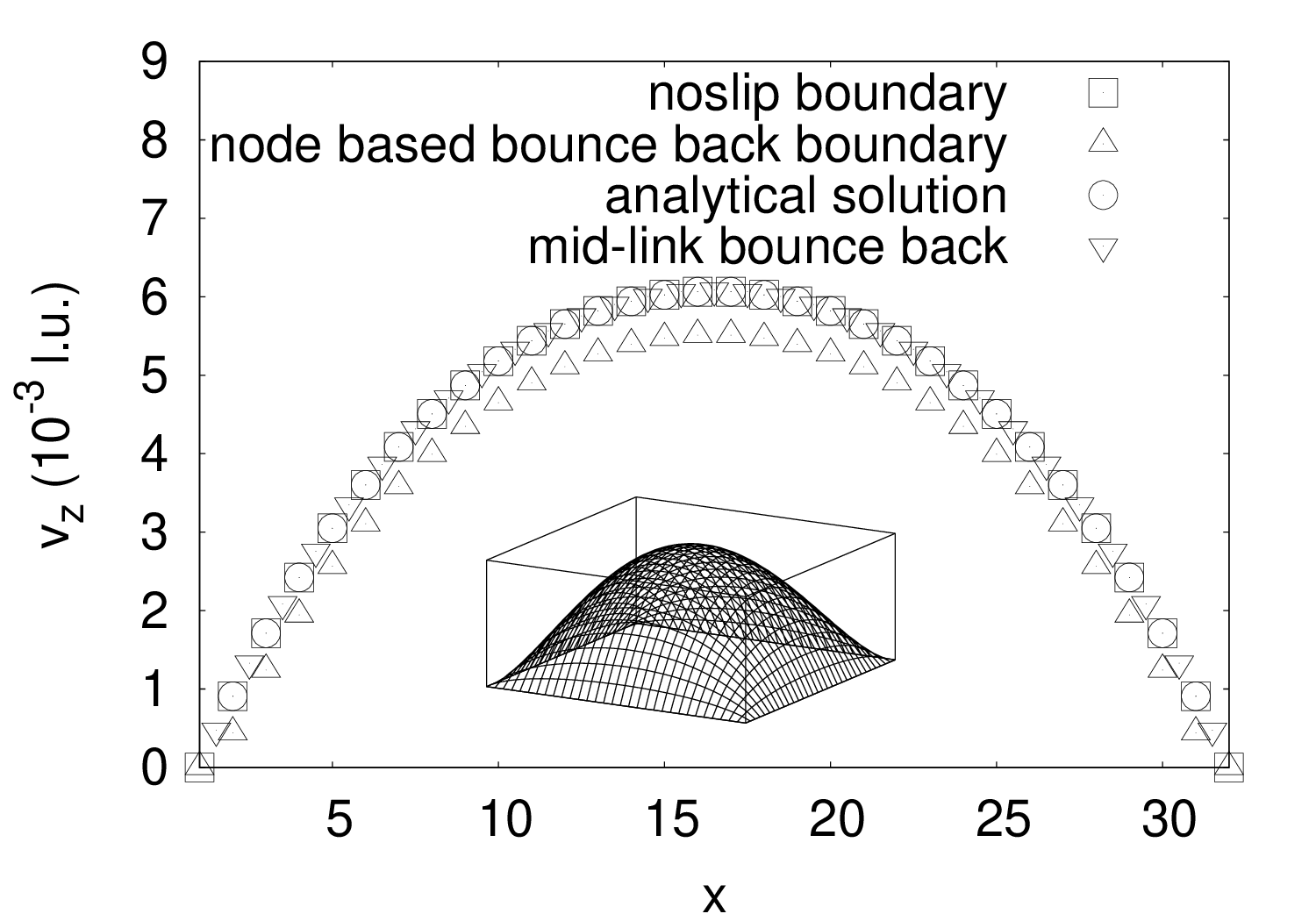}\\
  b)\hspace{-3ex}\includegraphics[width=0.85\linewidth]{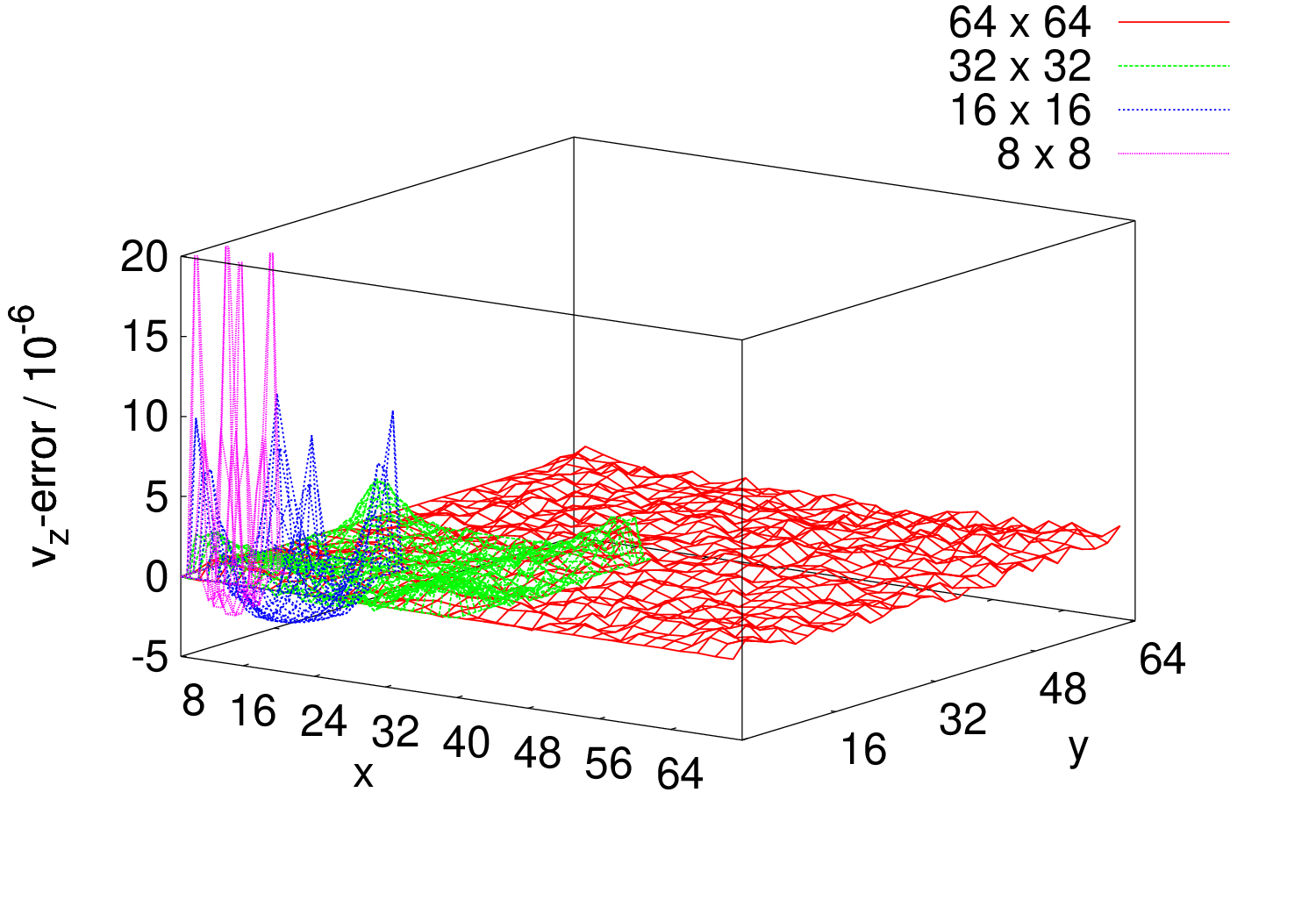}\\
  c)\hspace{-3ex}\includegraphics[width=0.8\linewidth]{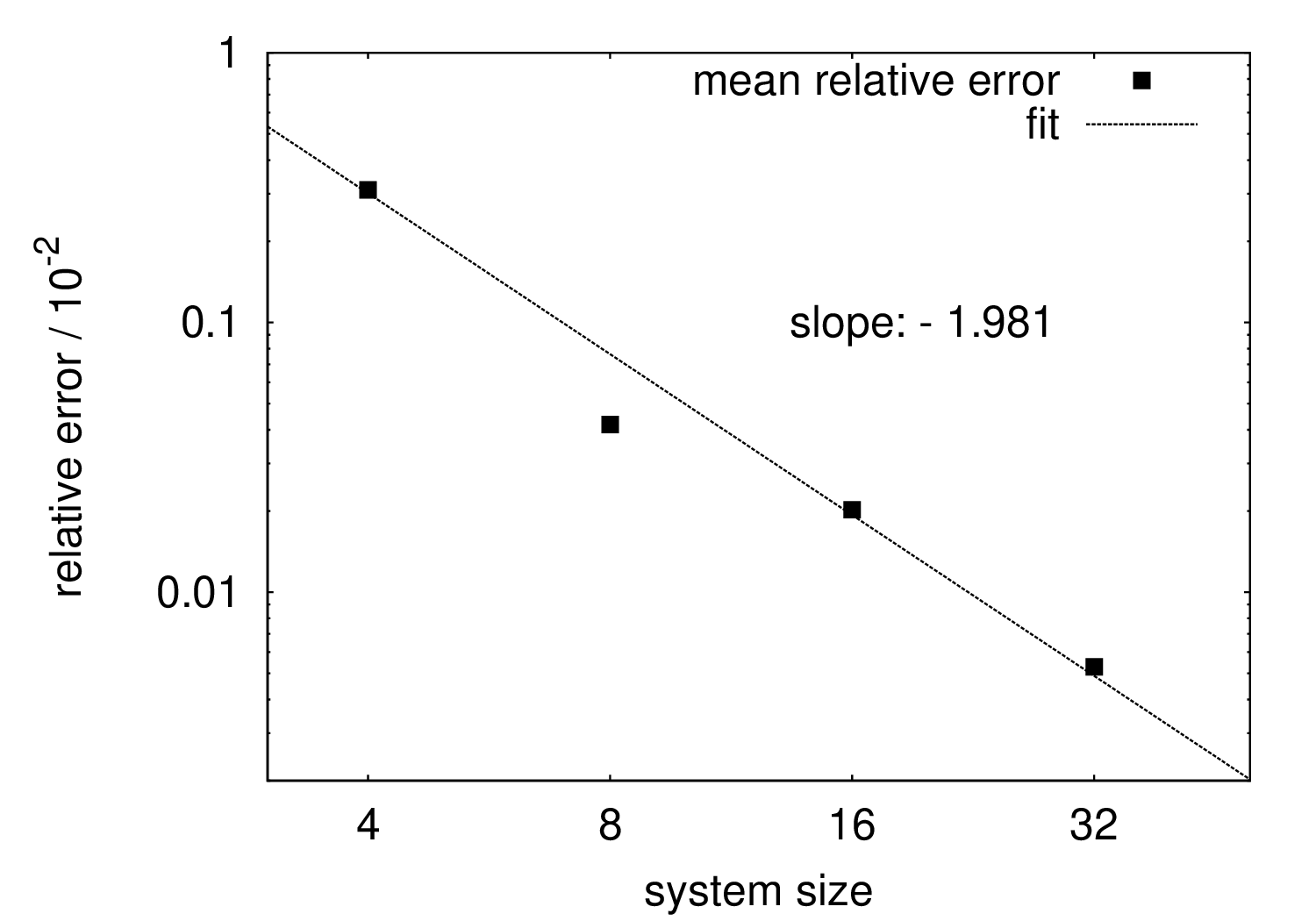}\\
  \caption{a) Velocity profile in a square channel averaged along the $yz$-planes for the noslip boundary
    condition (squares), for the node based bounce back rule (triangles), mid-link bounce back 
    (top down triangles), and the analytical solution. The noslip boundary condition collapses
    with the analytical solution, whereas the on-site bounce back boundary condition shows a kink 
    in the profile close to the boundary nodes. 
%This can be corrected by interpreting the position of 
%    the wall to be (depending on the BGK-relaxation time) somewhere 
%in between two nodes which restores second-order accuracy. 
The numerical results are obtained on a $32^3$ lattice and for
    the analytical solution the sum in \eq{eq:squarechannel} is truncated
after 50 terms. A 2D
    profile of the $z$-velocity in a $xy$-cross section is displayed as an inset. \\
    b) The relative error for system sizes between 8 and 64 nodes: in the corners next to the boundary
     the largest relative 
     deviations occur. Note that on the boundary nodes, where the truncation error of \eq{eq:squarechannel} 
     is largest, the deviation between simulation and approximated analytical solution is negligible. 
     Therefore, the deviation can be taken as a measure for the quality of the numerical result.\\
   c) The averaged error for different lattice resolutions versus the number of lattice nodes in each dimension confirming the second order accuracy of the no-slip boundary condition.}
  \label{fig_channelflow}
\end{figure}

Also in this second-order test we find that the numerical error is of the size of the 
floating point precision on the computer.
%We again find
%If we check the numerical error on each boundary node, we find errorbars of 
%$\pm 7.5\times 10^{-6}\%$ times the acceleration $\Delta \vec{v}$ due to the applied gravity.
%The numerical errors in this second-order test are of the same size as in the linear shear test.
This underlines that our boundary condition reproduces the velocity field up to second 
order. % and the remaining errors are due to the representation of floating point numbers on the computer.

Finally, we test our implementation by simulating flow through a square channel.
The analytical solution for the velocity of a pressure driven flow 
in a $b\times b$ square channel is\,\cite{Wieghardt57}
\bea
\vec{v}(x,y) &=& - 
\frac{\nabla p}{2\eta}\left[\frac{b^2}{4}-y^2-C(x)\right], \mathrm{with}\label{eq:squarechannel}\\
C(x)&=&\frac{8b^2}{\pi^3}\sum\limits_{n=0}^\infty (-1)^n
  \frac{\cosh\left(\frac{(2n+1)\pi x}{b}\right)\cos\left(\frac{(2n+1)\pi y}{b}\right)}{(2n+1)^3\cosh\left(\frac{(2n+1)\pi}{2}\right)}\nonumber
\eea
where $x \in [-b/2, b/2]$ and $y\in [-b/2,b/2]$ are the coordinates in the cross-section, with the
origin in the center of the channel. The pressure gradient $\nabla p$ is imposed by pressure boundaries 
and the dynamic viscosity is known from \eq{eq_visc}. The infinite sum in \eq{eq:squarechannel} can be truncated when a given accuracy is reached. We sum up 50 terms and 
compare this approximation to our numerical results on a $32^3$ domain, i.e., $b=15$ plus one layer 
of boundary nodes. In \fig{fig_channelflow}\,a) we compare the analytical solution from \eq{eq:squarechannel} with our simulation results. The velocity in $z$-direction is averaged over the $y$- and $z$-direction 
and the averaged value is plotted against the position in $x$-direction. A very good agreement of the
numerical result with the analytical solution can be seen. For comparison, results for the node based
bounce back rule are shown. For this boundary condition it is known that it is only first oderer accurate, 
which can be seen in the kink in the velocity profile close to the boundary nodes. It can however be 
made second order accurate by choosing the position of the wall somewhere (depending on the BGK 
relaxation time) in between two nodes, which is known as the mid-grid bounce back\,\cite{Succi01}. As one can see
in \fig{fig_channelflow}\,a), if the wall-position is chosen correctly for the bounce-back rule, 
a satisfiying accuracy can be achieved, too (top-down triangles). Note that the position
of the wall is shifted by half a lattice unit due to the different approach at the wall.

In \fig{fig_channelflow}\,b) the relative error depending on the size of the simulation is studied. 
There are three different errors involved. The truncation errors of the sum in \eq{eq:squarechannel}
can not be seen in this figure. If the sum is truncated after just a few terms, the error increases on
two of the edges. In the corners of the simulation domain the errors due to the discretization on 
the lattice remains. This is what determines the accuracy for relatively small simulations.
However, if the lattice is refined, this error decreases. It decreases with the square of the lattice
constant which is typical for second order accurate schemes. Another error which dominates for large 
systems is the floating point precision which is reflected in noisy data in the center of the 
simulation domain. This error is independent on the lattice constant.
In \fig{fig_channelflow}\,b) the relative error is shown for different system sizes. 
In the corners the error decreases with the system size, whereas the noise in the center is independent
by the system size. 
In  \fig{fig_channelflow}\,c) we plot the mean error averaged over the whole system against the number
of lattice nodes used for computation in each dimension. The slope of approximately $2$ confirms 
the second order accuracy of the boundary condition. For the simulations presented in \fig{fig_channelflow}
pressure boundaries according to \eq{eq_rhoin} are used and on the walls and edges we apply no-slip 
conditions as described in \sect{sect_noslip}. For this plot we use only the range in which the 
lattice size dependent error dominates. For 64 lattice nodes in each dimension, the floating point 
precision in one of our post processing steps dominates the overall error. Therefore, we only use
the smaller systems for this investigation.
The inset in \fig{fig_channelflow} shows the analytical velocity profile in a cross section perpendicular
to the extension of the square channel.

%==========================================================================================
\section{Conclusion}
%------------------------------------------------------------------------------------------
\noindent
We have derived an explicit local on-site flux boundary condition for LB
simulations on a D3Q19 lattice. Velocity terms up to second order enter
the derivation and this accuracy is also confirmed in the numerical tests.
The in- and outflux velocity underlies no restrictions to any peculiar
direction. We have demonstrated the numerical accuracy by comparing
simulation results for a flow through a tilted channel with the
theoretical expectation of a Poiseuille flow. Remaining errors can be
assigned to the discretization on the lattice and to rounding errors due
to the floating point representation. We have tested the boundary
condition in simulation of Poiseuille flow between two planar walls and in
shear flow. In those tests the simulation data fits exactly the analytical
solutions without any slip-parameter and independent on the BGK relaxation
time.  For this test we have used no-slip boundary conditions which are a
special case included in the general velocity boundary conditions.
Finally, we have tested the boundary condition by simulating the flow
through a square channel. The scaling of the numerical error with the
lattice resolution again confirms the second order accuracy.
%\begin{acknowledgement}
%\textbf{Acknowledgments.}
\section*{Acknowledgments}
\noindent
The German Research Foundation (DFG) is acknowledged for financial support (SFB 716 and EAMatWerk).
We thank A. Narv{\'a}ez, B. D{\"u}nweg, and U.~D. Schiller for fruitful discussions. 
%\end{acknowledgement}
\section*{Appendix}
\noindent
Here we give the expressions for the other boundaries not treated explicitly in the text. 
We start with the top-plane where we implement outflux for our simulations. We obtain
\bea
    \rho &=& \frac{1}{v_z+1}\left[f_{1} + f_{2} + f_{3} + f_{4} \right.\nonumber\\
         && + f_{7} + f_{11} + f_{12} + f_{8} + f_{19}  + \\
         &&    \left.   2( f_{5} + f_{9} + f_{13} + f_{15} + f_{17})\right] \nonumber
\eea
  or
\bea
    v_z &=& -1 + \frac{1}{\rho_0}\left[f_{1} + f_{2} + f_{3} + f_{4} \right.\nonumber\\
         && + f_{7} + f_{11} + f_{12} + f_{8} + f_{19} +  \\
         &&    \left.   2( f_{5} + f_{9} + f_{13} + f_{15} + f_{17})\right] \nonumber
\eea
with $v_z$ defined in positive $z$-direction. Here, the undetermined populations
after the streaming step are
\bea              
   f_{6} &=& f_{5} - \frac{1}{3}\rho v_z\,, \\
   f_{10} &=& f_{13} + \frac{\rho}{6} ( - v_z + v_x ) - N^z_x\,,\\
   f_{14}  &=& f_{9} + \frac{\rho}{6} ( - v_z - v_x ) + N^z_x\,,\\
   f_{16} &=& f_{17} + \frac{\rho}{6} ( - v_z + v_y ) - N^z_y\,,\\
   f_{18} &=& f_{15} + \frac{\rho}{6} ( - v_z - v_y ) + N^z_y
\eea
with $N^z_x$ and $N^z_y$ defined as previously, in \eq{eq_Nx} and \eq{eq_Ny}.
% For the components of the tensor ${\bf \Pi}$ we find
% \bea
%   \Pi_{11} &=& f_{1}+f_{2}+f_{7}+f_{8} +f_{11} +f_{12} \nonumber\\
%   &&+ 2(f_{9}+f_{13})-\frac{\rho v_z\!}{3}\,,\\
%   \Pi_{12} &=& \!\Pi_{21} = f_{7}-f_{8}+f_{12}-f_{11}\,,\\
%   \Pi_{13} &=& \!\Pi_{31} = 2(f_{9}\!-\!f_{13})-\frac{\rho v_x\!}{3} + 2 N^z_x\,,\\
%   \Pi_{22} &=& f_{3}+f_{4}+f_{7}+f_{8}+f_{11}+f_{12} \nonumber\\
%   &&+2(f_{15}+f_{17})-\frac{\rho v_z\!}{3}\,,\\
%   \Pi_{23} &=& \!\Pi_{32} = 2(f_{15}\!-\!f_{17})-\frac{\rho v_y\!}{3} + 2 N^z_y\,,\\
%   \Pi_{33} &=&  2 (f_{5} + f_{9} + f_{13} + f_{15} + f_{17}) \nonumber\\
%   && -\rho v_z \,.
% \eea

For the left, right, front and back boundaries, which we do not use in this work one finds
the following expressions. For the left ($x=0$) boundary as inlet,
\bea
    \rho &=& \frac{1}{1-v_x}\left[ f_{3} + f_{4} + f_{5} + f_{6} \right.\nonumber\\
    &+&  f_{15} + f_{16} + f_{17} + f_{18} + f_{19} \label{eq_rhooutx} \\
    &+&\left.  2 ( f_{2} + f_{11} + f_{12} + f_{13} + f_{14})\right]\,,\nonumber
\eea                                      
or
\bea
    v_x &=& 1 - \frac{1}{\rho_0}\left[ f_{3} + f_{4} + f_{5} + f_{6} \right.\nonumber\\
    &+&  f_{15} + f_{16} + f_{17} + f_{18} + f_{19} \label{eq_voutx} \\
    &+&\left.  2 ( f_{2} + f_{11} + f_{12} + f_{13} + f_{14})\right]\,,\nonumber
\eea                                      
and
\bea              
   f_{1} &=& f_{2} + \frac{1}{3}\rho v_x \,,\\
   f_{8} &=& f_{11} + \frac{\rho}{6} ( v_x - v_y ) + N^x_y \,,\\
   f_{7} &=& f_{12} + \frac{\rho}{6} ( v_x + v_y ) - N^x_y \,,\\
   f_{9}  &=& f_{14} + \frac{\rho}{6} ( v_x + v_z) - N^x_z \,,\\
   f_{10} &=& f_{13} + \frac{\rho}{6} ( v_x - v_z) + N^x_z
\eea
with
\bea
  N^x_y &=&  \frac{1}{2}\left[f_{3}+f_{15}+f_{16}-(f_{4}+f_{17}+f_{18})\right] \nonumber\\
      &&   - \frac{1}{3}\rho v_y \,,\label{eq_Nxy}\\
  N^x_z &=&  \frac{1}{2}\left[f_{5}+f_{17}+f_{15}-(f_{6}+f_{16}+f_{18})\right] \nonumber\\
      &&   - \frac{1}{3}\rho v_z \label{eq_Nxz}\,.
\eea

At the right boundary (outlet) we have 

\bea
    \rho &=& \frac{1}{v_x+1}\left[ f_{3} + f_{4} + f_{5} + f_{6} \right.\nonumber\\
    &+&  f_{15} + f_{16} + f_{17} + f_{18} + f_{19} \label{eq_rhoinx} \\
    &+&\left.  2 ( f_{1} + f_{7} + f_{8} + f_{9} + f_{10})\right]\,,\nonumber
\eea                                      
or
\bea
    v_x = -1 &+& \frac{1}{\rho_0}\left[ f_{3} + f_{4} + f_{5} + f_{6} \right.\nonumber\\
    &+&  f_{15} + f_{16} + f_{17} + f_{18} + f_{19} \label{eq_uinx} \\
    &+&\left.  2 ( f_{1} + f_{7} + f_{8} + f_{9} + f_{10})\right]\,,\nonumber
\eea                                      
and
\bea              
   f_{2} &=& f_{1} - \frac{1}{3}\rho v_x \,,\\
   f_{11} &=& f_{8} + \frac{\rho}{6} ( - v_x + v_y ) - N^x_y \,,\\
   f_{12} &=& f_{7} + \frac{\rho}{6} ( - v_x - v_y ) + N^x_y \,,\\
   f_{14}  &=& f_{9} + \frac{\rho}{6} ( - v_x - v_z) + N^x_z \,,\\
   f_{13} &=& f_{10} + \frac{\rho}{6} ( - v_x + v_z) - N^x_z \,.
\eea

At the front ($y=0$) boundary as inlet, one finds
\bea
    \rho &=& \frac{1}{1-v_y}\left[ f_{1} + f_{2} + f_{5} + f_{6} \right.\nonumber\\
    &+&  f_{9} + f_{10} + f_{13} + f_{14} + f_{19} \label{eq_rhoiny} \\
    &+&\left.  2 (f_{4} + f_{8} + f_{12} + f_{17} + f_{18})\right]\,,\nonumber
\eea                                      
or
\bea
    v_y = 1 &-& \frac{1}{\rho_0}\left[ f_{1} + f_{2} + f_{5} + f_{6} \right.\nonumber\\
    &+&  f_{9} + f_{10} + f_{13} + f_{14} + f_{19} \label{eq_uiny} \\
    &+&\left.  2 (f_{4} + f_{8} + f_{12} + f_{17} + f_{18})\right]\,,\nonumber
\eea                                      
and
\bea              
   f_{3}  &=& f_{4}  + \frac{1}{3}\rho v_y \,,\\
   f_{7}  &=& f_{12} + \frac{\rho}{6} ( v_y + v_x ) - N^y_x \,,\\
   f_{11} &=& f_{8}  + \frac{\rho}{6} ( v_y - v_x ) + N^y_x \,,\\
   f_{15} &=& f_{18} + \frac{\rho}{6} ( v_y + v_z ) - N^y_z \,,\\
   f_{16} &=& f_{17} + \frac{\rho}{6} ( v_y - v_z ) + N^y_z
\eea
with
\bea
  N^y_x &=&  \frac{1}{2}\left[f_{1}+f_{9}+f_{10}-(f_{2}+f_{13}+f_{14})\right] \nonumber\\
      &&   - \frac{1}{3}\rho v_x \,,\label{eq_Nyx}\\
  N^y_z &=&  \frac{1}{2}\left[f_{5}+f_{9}+f_{13}-(f_{6}+f_{10}+f_{14})\right] \nonumber\\
      &&   - \frac{1}{3}\rho v_z \label{eq_Nyz}\,.
\eea

At the back (outlet) the density is given by
\bea
    \rho &=& \frac{1}{v_y+1}\left[ f_{1} + f_{2} + f_{5} + f_{6} \right.\nonumber\\
    &+&  f_{9} + f_{10} + f_{13} + f_{14} + f_{19} \label{eq_rhoouty} \\
    &+&\left.  2 (f_{3} + f_{7} + f_{11} + f_{15} + f_{16})\right]\,,\nonumber
\eea      
%\vspace{1pt}
or the velocity reads
\bea
    v_x &=& -1 + \frac{1}{\rho_0}\left[ f_{1} + f_{2} + f_{5} + f_{6} \right.\nonumber\\
    &+&  f_{9} + f_{10} + f_{13} + f_{14} + f_{19} \label{eq_uouty} \\
    &+&\left.  2 (f_{3} + f_{7} + f_{11} + f_{15} + f_{16})\right]\,,\nonumber
\eea     
and the distributions are
\bea              
   f_{4}  &=& f_{3}  - \frac{1}{3}\rho v_y \,,\\
   f_{12} &=& f_{7}  + \frac{\rho}{6} ( - v_y - v_x ) + N^y_x \,,\\
   f_{8}  &=& f_{11} + \frac{\rho}{6} ( - v_y + v_x ) - N^y_x \,,\\
   f_{18} &=& f_{15} + \frac{\rho}{6} ( - v_y - v_z ) + N^y_z \,,\\
   f_{17} &=& f_{16} + \frac{\rho}{6} ( - v_y + v_z ) - N^y_z \,.
\eea

% \bibliography{hecht}
\bibliography{lbpaper}

\end{document}